\documentclass[10 pt,final,journal,twocolumn,twoside,letterpaper]{IEEEtran}
\usepackage{amssymb,amsmath,epsfig}
\usepackage{booktabs,multirow,comment}
\usepackage{slashbox,bigstrut,color}

\begin{document}

\title{Pushing the Communication Speed Limit of a Noninvasive BCI Speller}
\author{Po~T.~Wang, Christine~E.~King, An~H.~Do, and Zoran~Nenadic%
\thanks{P.~T.~Wang is with the Department of Biomedical Engineering, University of California, Irvine, CA 92697, USA}
\thanks{C.~E.~King is with the Department of Biomedical Engineering, University of California, Irvine, CA 92697, USA}
\thanks{A.~H.~Do is with the Department of Neurology, University of California, Irvine, CA 92697, USA}
\thanks{Z.~Nenadic is with the Department of Biomedical Engineering and the Department of Electrical Engineering and Computer Science, University of California, Irvine, CA 92697, USA}}
\maketitle 
\begin{abstract}
Electroencephalogram (EEG) based brain-computer interfaces (BCI) may provide a means of communication for those affected by 
severe paralysis. However, the relatively low information transfer rates (ITR) of these systems, currently limited to 1 bit/sec, present a serious obstacle to their widespread adoption in both clinical and non-clinical applications. Here, we report on the development of a novel noninvasive BCI communication system that achieves ITRs that are severalfold higher than those previously reported with similar systems. Using only 8 EEG channels, 6 healthy subjects with little to no prior BCI experience 
selected characters from a virtual keyboard with sustained, error-free, online ITRs in excess of 3 bit/sec. By factoring in the time spent to notify the subjects of their selection, 
practical, error-free typing rates as high as 12.75 character/min were achieved,    
which allowed subjects to correctly type a 44-character sentence in less than 3.5 minutes. We hypothesize that ITRs can be further improved by optimizing the parameters of the interface, while practical typing rates can be significantly improved by 
shortening the selection notification 
time. 
These results provide compelling evidence that the ITR limit of noninvasive BCIs has not yet been reached and that further investigation into this matter is both justified and necessary. 
\end{abstract}
\begin{keywords} Brain-computer interfaces, P300 speller, information transfer rate.
\end{keywords}
\section{Introduction}
Noninvasive electroencephalogram (EEG) based brain-computer interfaces (BCIs) may provide a means of communication for those affected by locked-in syndrome~\cite{nbirbaumer:99} or other forms of severe paralysis. These systems rely on predictable EEG patterns that are 
translated into control signals for real-time operation of external devices~\cite{Wolpaw2002}. Thus, individuals with severe paralysis may benefit from BCI technology by 
bypassing the disrupted motor pathways and operating prostheses directly from the brain. Paralyzed individuals have successfully used EEG-based BCIs to operate computer cursors~\cite{jrwolpaw:04,djmcfarland:10}, a virtual reality wheelchair~\cite{rleeb:07}, a virtual reality avatar~\cite{ptwang:10a}, and  functional electrical stimulation devices~\cite{ahdo:11, grmuellerputz:05, gpfurtscheller:03}.

One of the most robust 
EEG-based BCI communication systems is the P300 speller, developed originally by Farwell and Donchin~\cite{Farwell1988}. This communication protocol relies on the P300 evoked potential~\cite{ssutton:65}---a positive deflection in EEG signals, observed predominantly over the parietal lobe, which occurs $\sim$300 msec after the presentation of an infrequent, task-relevant stimulus. In a large majority of P300 spelling systems, the presentation of task-relevant stimuli conforms to the 
visual oddball paradigm~\cite{kcsquires:77}, where the subject is instructed to pay attention to a rare stimulus in a random sequence of 
stimuli presented on a computer screen. The subject's intentions can then be decoded in real time by 
detecting the presence of the P300 potential that coincides with the illumination of letters from a virtual keyboard. 
It was hypothesized in~\cite{Farwell1988} that such a system could achieve information transfer rates (ITRs) as high as 0.2 bit/sec (or 2.3 characters/min). 
  
While subsequent studies (e.g.~\cite{djkrusienski:08,gtownsend:10}) have managed to optimize the original BCI spelling system and thus significantly improve its performance, the achieved ITRs are still relatively modest and fall well below those of communication and/or control systems that rely on residual motor function, such as eye movements~\cite{tehutchinson:89}. Whether used in spelling, computer cursor movement, or other applications,  
 it is generally accepted that the ITR limit of EEG-based BCIs is $\sim$1 bit/sec~\cite{Wolpaw2002}, which remains a major obstacle to their adoption in both clinical and non-clinical applications. 
 
 In this article, we report on the development of a novel EEG-based BCI communication protocol, where subjects with little to no prior BCI experience were able to achieve sustained, error-free ITRs in excess of $3$ bit/sec in a real-time typing test. These bit rates are severalfold higher than those previously reported by P300 spellers and EEG-based BCIs. We hypothesize that optimization of the interface parameters and user training may further increase the communication speed limit of these systems, which may have a significant impact on their adoption in both clinical and non-clinical applications.     

\section{Methods}
\subsection{Study Protocol}\label{sec:sp1}
Six able-bodied individuals (see Table~\ref{tab:demographics} for demographic data) were recruited for this study. All subjects had normal or normal corrected vision and no cognitive or neurological impairments. 
The study  was approved by the University of California Irvine Institutional
Review Board.

\begin{table}[!htbp]
\caption{Demographic data of the study participants.}\label{tab:demographics}
\centering
\begin{tabular}{cccll}\toprule
{Subject} & Gender & Age & Prior BCI experience & Native speaker\\ \midrule
A & F & 23 & Yes (3 hours) & Yes \\
B & M & 40 & Yes (10 hours) & No \\
C & M & 29 & Yes (1 hour) & Yes \\
D & M & 22 & No  & Yes \\
E & M & 24 & No & Yes \\ 
F & F & 56 & Yes (10 hours) & Yes \\ \bottomrule
\end{tabular}
\end{table}

Each participant completed three experimental sessions performed on three different days over the course of 1--3 weeks. Within each daily session, a subject performed BCI spelling experiments at three different interface speeds (see Table~\ref{tab:sp}). 
For each speed, a short training procedure was performed, followed by 1--3 online spelling sessions. A detailed description of these procedures is given in Section~\ref{sec:ep}.

\begin{table}[!htbp]
\caption{Breakdown of the study protocol.}\label{tab:sp}
\centering
\begin{tabular}{clll}\toprule
{Day} & \multicolumn{3}{c}{Interface speed} \\ \midrule
1 & slow & medium & fast\\
2 & medium & fast & slow \\
3 & fast & slow & medium \\ \bottomrule
\end{tabular}
\end{table}

Online sessions were performed in a free spelling mode~\cite{nbirbaumer:99}. Specifically, the subjects were asked to {\it correctly} spell the following 44-character sentence: {\tt THE QUICK BROWN FOX JUMPS OVER THE LAZY DOG*}. This includes spaces and the symbol {\tt *} at the end, which serves to exit the interface. 
Note that each letter of the English alphabet appears at least once in the sentence. 
In the case of a typing error, the subjects used backspace to delete erroneously selected characters, and then proceeded with the correct sequence of letters. All subjects were able to 
type the benchmark sentence free of errors and exit the interface (see Section~\ref{sec:r} for details). Also, the subjects had no trouble memorizing the sentence and were able to track their spelling progress on the computer screen. The total daily involvement per subject 
was 2--3 hours. 
 
\subsection{Data Acquisition}
\begin{figure}[!hbpt]
        \centering
                \includegraphics[width=1.0\columnwidth]{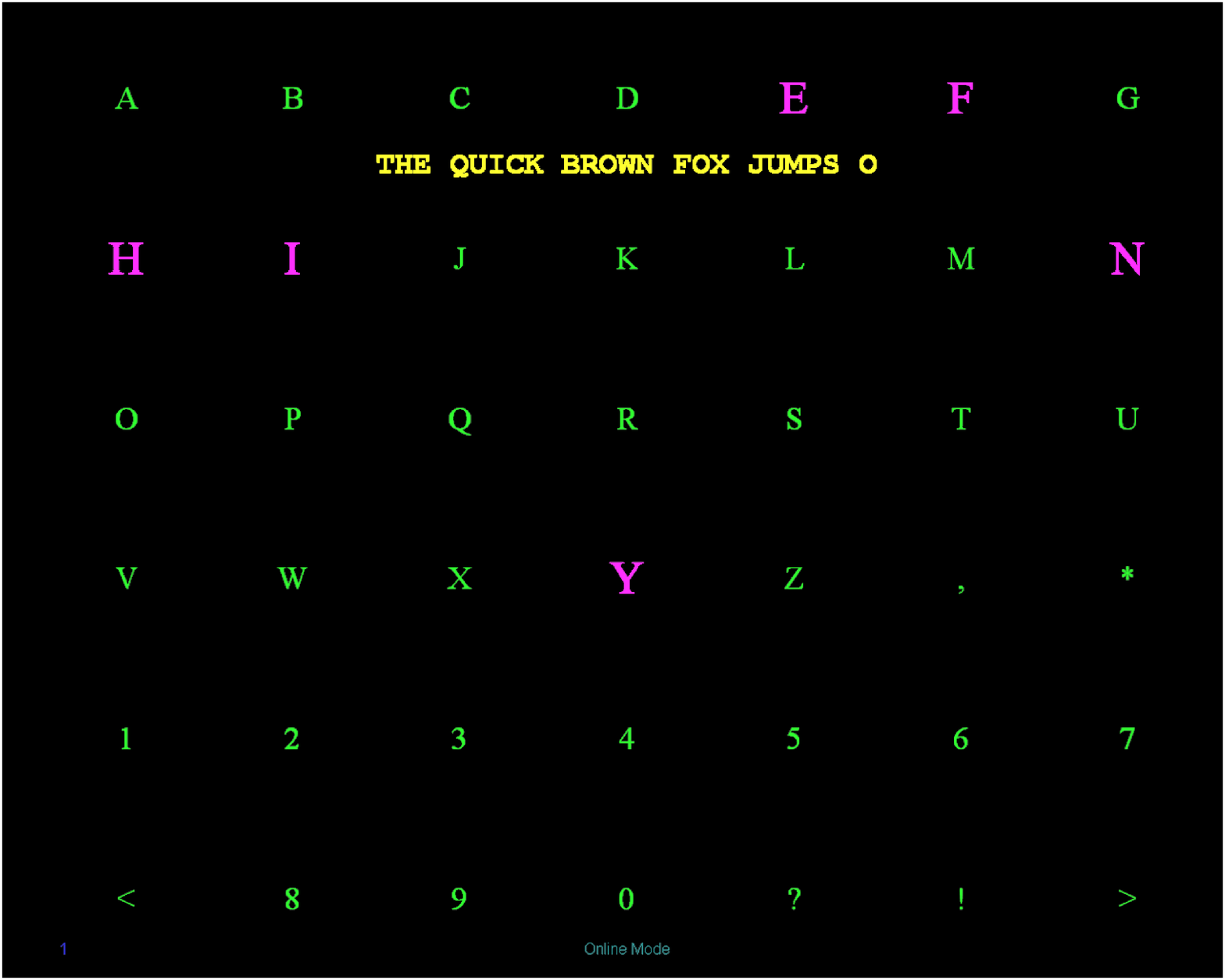}
                \caption{A screenshot of the character matrix. The illuminated 
                letters are bold-faced and highlighted (pink).
        The prompt (yellow) shows the typing progress.}
        \label{fig:screenshot}
\end{figure}

Each subject was seated $\sim$0.9 m from a computer monitor that displayed a 6$\times$7 matrix of characters (see Fig.~\ref{fig:screenshot}). An EEG cap (Compumedics USA, Charlotte, NC) with 19 sintered Ag-AgCl electrodes, arranged according to the 10-20 International Standard, was used for EEG recording. Conductive gel was applied to a subset of eight electrodes at the following locations: C3, Cz, C4, P3, Pz, P4, O1, and O2 (see Fig.~\ref{fig:erp}). 
Ear clip electrodes, A1 and A2, were linked and used as a reference electrode. However, if the 30-Hz ear-to-ear impedance was above 10 k$\Omega$, only the ear with the lower impedance was used as a reference. The impedances between the reference and each of the eight electrodes were $<$5 k$\Omega$. The signals were then amplified (gain: 5,000) and band-pass filtered (1--35 Hz) using 8 single-channel EEG bioamplifiers (Biopac Systems, Goleta, CA), and were digitized (sampling rate: 200 Hz, resolution: 16 bits) by the MP150 acquisition system (Biopac Systems). The data acquisition and experimental protocols were controlled by 
custom-made Matlab (Mathworks, Natick, MA) scripts. EEG data recorded during training procedures were 
analyzed offline, while those recorded during online sessions were analyzed in real-time (details in Section~\ref{sec:ep}).  
\subsection{Experimental Protocol}\label{sec:ep}       
\begin{figure}[!htbp]
        \centering
                \includegraphics[width=3.25in]{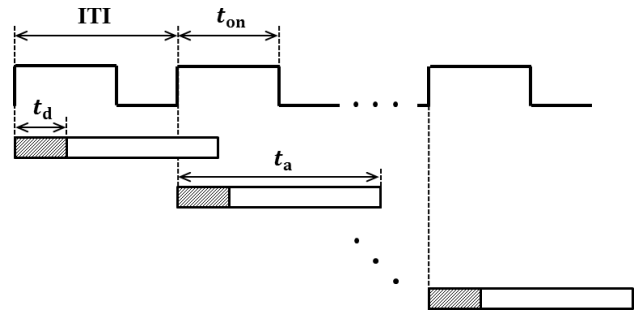}
        \caption{The timing diagram of the experimental protocol.}
        \label{fig:exp_protocol}
\end{figure}

For each choice of interface speed (see Table~\ref{tab:sp}), a training session was performed first followed by an online spelling test. A training session started by instructing the subject to focus on a specific character 
for 30 sec. Within this time frame, the characters were illuminated randomly in groups of 6 in a block randomized fashion, i.e. after a single cycle consisting of seven illuminations, all 42 characters have been illuminated exactly once. The details of the randomization algorithm are described in Appendix~\ref{a:groc}. Its main function is to group characters according 
to the frequency in which they appear in the English language so that groups containing more frequent letters can be illuminated earlier in the cycle. The benefits of this 
algorithm 
will later be elaborated upon. 
Also, note that as only one of the groups contains the desired character, the ratio of oddball and non-oddball stimuli in the training sessions is 1:6. Upon completing the cycle, 
the groups were re-randomized, and the whole procedure was repeated for a total of 30 sec. The frequency of illumination was controlled by the inter-trial interval (ITI) (see Fig.~\ref{fig:exp_protocol}), with a duty cycle of 60\% ($t_{\mathrm{on}}/\mathrm{ITI}=0.6$). After 30 sec, a short break ensued, during which the subject was instructed to focus on another character, and the whole procedure was repeated for a total of 10 characters. Finally, the whole training session including breaks lasted $\sim$6.5 min. 

In response to each illumination, $t_{\mathrm{a}}=400$ msec of EEG data were acquired and stored for offline analysis. 
Throughout this article, a single 400-msec data segment is referred to as a trial. The total number of trials in each training procedure depended on the choice of ITI. 
In particular, the slow, medium and fast interface speeds (Table~\ref{tab:sp}) corresponded to an ITI of 400, 240 and 160 msec, respectively.
Note that for the medium and fast speeds, the successive trials partially overlapped (as illustrated in Fig.~\ref{fig:exp_protocol}).   

During online sessions, the 
illumination of characters was controlled as follows. In the initial stage (referred to as stage 1), the computer illuminated characters in groups of six (see Fig.~\ref{fig:screenshot}) in the same manner as done in training sessions. In response to each illumination, an EEG trial 
was processed in real time and classified (see Section~\ref{sec:spac}). As long as the trials were classified as non-oddball, the interface kept cycling through stage 1. 
Once an oddball trial was detected, the BCI computer transitioned to stage 2, where individual characters from the selected group were illuminated. Note that faster transition to stage 2 was facilitated by illuminating the more frequent characters earlier in the stage-1 cycle. 

Similar to stage 1, the illumination order in 
stage 2 was based on the character's relative frequencies (see Appendix~\ref{a:groc}). Once an individual character was selected (by classifying its corresponding trial as oddball), the interface highlighted 
the selected character using a green rectangular background and paused for 3 sec to let the subject know of the decision. In addition, the selected character was copied to the typing prompt (see Fig.~\ref{fig:screenshot}) so that 
 the typing progress could be tracked. The BCI computer then transitioned back to stage 1, and the whole process was repeated. An exception to this rule occurred when the BCI computer found a limited number ($\le$5) 
of dictionary completions to the partially spelled word. In this case, the completion characters were illuminated individually in a random order. Repeated failure (over three cycles) to select a character in this single-character illumination stage  resulted in 
the BCI computer defaulting to stage 1. 

 In addition to the aforementioned single-trial selection method, the BCI 
 kept track of the highest letter count. More specifically, 
   this method integrates the evidence, expressed as the posterior probability (see Section~\ref{sec:spac}) of individual characters, to find  
a character that has had the highest integrated probability for over 10 consecutive trials 
to immediately classify and select it as oddball, thus bypassing stage 2. 
This mechanism is useful when the evidence based on a single trial is not sufficiently strong to warrant the character selection. 
Upon character selection, the counter is reset to zero and all the posterior probabilities are reset to 1/42.

\subsection{Signal Processing and Classification}\label{sec:spac}
Within each trial (see Fig.~\ref{fig:exp_protocol}), the first $t_\mathrm{d}=100$ msec 
of data was presumed to contain no useful information due to the lag in visual information processing~\cite{sthorpe:96}, and so the trials were effectively truncated to 300 msec. 
Subsequently, 
trials from the training procedure, represented by $8\times60$ matrices (8 channels, $0.3\times200$ samples), were 
reshaped 
into 480-dimensional (480-D) vectors. To allow accurate estimation of data statistics under both oddball and non-oddball conditions, and to facilitate subsequent classification of trials into these two classes, the dimension of data was reduced using a combination of  
classwise principal component analysis (CPCA)~\cite{kdas:09, Das2007} and approximate information discriminant analysis (AIDA)~\cite{Das2008}. 

For binary pattern recognition 
problems, 
CPCA projects high-dimensional data onto a pair of subspaces locally adapted to individual classes. Due to its nonlinear (piecewise linear) nature, CPCA is well-suited for 
problems where 
traditional linear dimensionality reduction techniques may be inadequate. In addition, unlike PCA and other nonlinear dimensionality reduction methods~\cite{stroweis:00}, CPCA is a supervised learning technique, and it therefore takes advantage of the known class information. 
In the present study, implementing CPCA with default parameters~\cite{kdas:09} typically resulted in dimension reduction from 480 to 20--30. 
A detailed description of this technique 
can be found in~\cite{kdas:09}. 

To enhance class separability while further reducing the dimension of data, AIDA~\cite{Das2008} was used. AIDA 
represents an approximation of an information-theoretic technique~\cite{znenadic:07}, which 
extracts features by maximizing the mutual information between the class labels and data. However, unlike computationally expensive information-theoretic methods~\cite{znenadic:07,ktorkkola:03}, AIDA retains the computational simplicity characteristics of linear, second-order techniques, such as 
linear discriminant analysis~\cite{roduda:01}. Specifically, the feature extraction matrix, $T_{\mathrm{AIDA}}$, is found through eigen-decomposition. In this study, 1-D features were extracted 
in this manner, i.e. $T_{\mathrm{AIDA}}\in \mathbb{R}^{m\times 1}$, where $m$ is the subspace dimension found by CPCA ($m=$ 20--30). A detailed account of AIDA can be found in~\cite{Das2008}. 
 
 Once 1-D features were extracted, 
 a linear Bayesian classifier was designed in the feature domain: 
\begin{equation}
\label{eq:bayes}
\frac{p(o\,|\,F^{\star})}{p(e\,|\,F^{\star})}
\begin{array}{c}
o\\[-0.075in]
>\\[-0.075in]
<\\[-0.075in]
e
\end{array}\theta
\end{equation} 
where 
$p(o\,|\,F^{\star})$ and $p(e\,|\,F^{\star})$ are the posterior probabilities of oddball and non-oddball classes given the observed feature $F^{\star}$, respectively. The 
equation~(\ref{eq:bayes}) reads: ``classify $F^{\star}$ as oddball if $p(o\,|\,F^{\star})/p(e\,|\,F^{\star})>\theta$, and vice versa.'' The threshold $\theta=\lambda_{\mathrm{FA}}/\lambda_{\mathrm{OM}}$ represents the 
false alarm in which case the classifier~(\ref{eq:bayes}) minimizes the total risk function~\cite{roduda:01,znenadic:05a} (see Appendix~\ref{a:cd} for details).

The performance of the above feature extraction and classification methods was tested on the training-procedure data using stratified 10-fold cross-validation (CV)~\cite{rkohavi:95}. Briefly, the EEG trials were randomly separated into 10 groups (folds) while approximately preserving the 1:6 oddball to non-oddball ratio in each group. The data from 9 folds were then used to train the parameters of CPCA, AIDA, and the Bayesian classifier~(\ref{eq:bayes}), and the data from the remaining fold were transformed into the feature domain and classified assuming an equal cost of omissions and false alarms (i.e. $\theta = 1$). The procedure was then repeated until all 10 folds were exhausted, each time designating a different fold for classification. The number of misclassified trials  was  used to calculate the probabilities of omission and false alarm errors. To estimate the standard deviation on these errors,  the 10-fold CV procedure was repeated 10 times, each time re-randomizing the grouping of trials into folds. The parameters of CPCA, AIDA, and the Bayesian classifier were then saved for the online procedure. 
    
In the online sessions, real-time acquired trials were reshaped in the same way as the training data, and subsequently transformed into features using CPCA and AIDA transformations. The 1-D features were then classified with the Bayesian classifier~(\ref{eq:bayes}). For stage 1 online classification, the same threshold, $\theta=1$, was used as in the training session. For stage 2 online classification, or when the interface was in a single-character illumination mode, the threshold changed to $\theta=0.5$. This choice reflects our empirical finding that P300 weakens when characters are illuminated individually. 

\subsection{Information Transfer Rate Calculations}\label{sec:itrc}
The described BCI system can be modeled as a binary communication channel (see Appendix~\ref{a:itrc}) whose inputs are user intentions and outputs are the decoded intentions. The amount of information per transmission is given by the mutual information between inputs and outputs~\cite{tmcover:91}, i.e. 
\begin{equation}
\label{eq:mi}
I(\mathrm{in},\mathrm{out}) = H(\mathrm{out}) - H(\mathrm{out}|\mathrm{in})
\end{equation} 
which measures the reduction in the output uncertainty by providing the input. 
The explicit formula for calculating~(\ref{eq:mi}) is given in Appendix~\ref{a:itrc}.
The ITR can then be calculated as: $\mathrm{ITR}=B\,I(\mathrm{in},\mathrm{out})$, where $B$ is the number of transmissions (character illuminations) per unit of time.  

For online performances, the total time $T$ to correctly type the benchmark sentence (44 characters) was recorded by the BCI computer. This time included the 3 sec pause after each selection that allowed subjects to be notified of their selection, track the typing progress, and visually locate the next desired character. This was true regardless of whether a correct or incorrect selection was made. In addition, the subjects were required to correct the incorrect selections by backspacing. While in this case, the selection of ${\tt <}$ (backspace) represents an intended action, backspaces were \textit{not} counted as correct selections since their purpose is to merely rectify previously committed error(s). As stringent as these requirements are, we believe that they set a standard for the definition of ITR that is completely immune to bit rate manipulations. More formally, practical error-free ITR is defined as:
\begin{equation}
\label{eq:itr}
\mathrm{ITR} = \frac{N_c}{T}\log_2{|\mathcal{A}|}
\end{equation}         
where $N_c$ is the number of correctly typed characters and $|\mathcal{A}|$ is the size of the alphabet ($N_c=44$, $|\mathcal{A}|=42$ in this study).
 
\section{Results}\label{sec:r}
\begin{figure}[!htbp]
        \centering
                \includegraphics[width=2.75in]{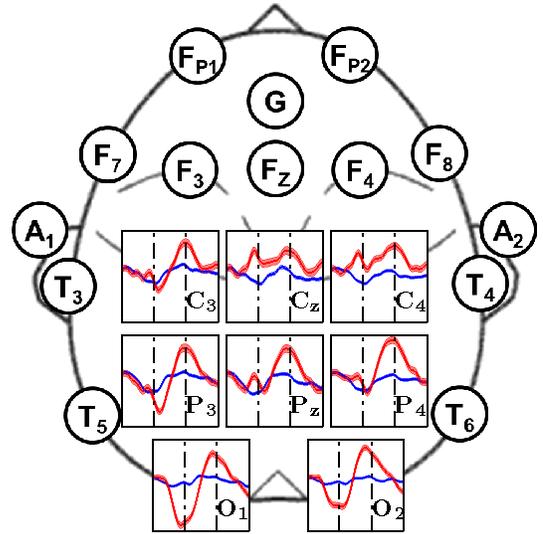}
        \caption{Event-related potentials of oddball (red) and non-oddball (blue) trials for Subject B, collected at the slow interface speed. The error bars represent the standard error of mean. Each panel is 18 $\mu$V $\times$ 300 msec, with the grid lines corresponding to 200 and 300 msec post-stimulus.} 
        \label{fig:erp}
\end{figure}

The data from each training procedure were used for offline estimation of feature extraction and classification parameters, as outlined in Section~\ref{sec:spac}. Event-related potential (ERP) analysis (obtained by averaging oddball and non-oddball trials), consistently revealed that subjects utilized both N200, mostly visible on the occipital lobe $\sim$190 msec post-stimulus (see Fig.~\ref{fig:erp}), and P300 which was present on all channels $\sim$285-300 msec post-stimulus. 
This is consistent with findings reported by other groups, e.g.~\cite{djkrusienski:08,ewsellers:07}. In addition, cross-correlation analysis revealed that the prominent positive potential seen in the fronto-parietal areas (most notably at electrodes Cz and C4) approximately 190 msec  post-stimulus was likely coming from the N200 source, albeit recorded from the opposite side of the dipole, and hence the phase reversal~\cite{plnunez:06}.   

The offline performances estimated through 10-fold CV and expressed as the probability of correct classification are presented in Table~\ref{tab:offline}.  Classification rates as high as 97.4\% were achieved, and all subjects performed significantly above the 85.71\% threshold, determined 
to be the chance-level performance of the Bayesian classifier (see Appendix~\ref{a:itrc}). 
The number of 
trials varied depending on the ITI: 
750 (ITI=400 msec), 
1250 (ITI=240 msec), and 1870 (ITI=160 msec). To rule out the effect of the sample size on the achieved classifier performances, the feature extraction and classifier training procedures were repeated in the case of ITI=250 msec and ITI=160 msec by randomly sub-selecting 750 trials while preserving the 1:6 oddball-to-non-oddball ratio. 
The classification rates were not significantly different than those using all available trials, so the differences in the offline performances observed across ITIs are likely caused by other factors, such as the dependence of the P300 amplitude on ITI time~\cite{clgonsalvez:02}. Finally, the offline ITRs were calculated using Eq.~(\ref{eq:mi}).
Based on  these performances, all subjects were expected to have purposeful control of the BCI in the online mode (see below).

\begin{table}[!htbp]
\caption{Offline performances of subjects as assessed through 10-fold CV. Rows show mean performance (top), best performance (middle), and ITR of best performance (bottom).}\label{tab:offline}
\centering
\begin{tabular}{clll}\toprule
\multirow{3}{*}{Subject} & \multicolumn{3}{c}{ITI (msec)}  \\ 
 & 400 & 240 & 160 \\ \midrule
\multirow{1}{*}{A} & 95.7$\pm$1.5\% & 94.1$\pm$1.1\% & 93.4$\pm$1.6\% \\ 
 & 97.0\%& 94.9\%& 94.4\% \\ 
 & 0.403$^{\star}$ & 0.308 & 0.293 \\ \midrule
 \multirow{1}{*}{B} & 95.2$\pm$1.9\% & 93.9$\pm$2.5\% & 94.3$\pm$2.9\% \\
 & 97.4\%& 96.2\%& 96.6\% \\ 
  & 0.427$^{\star\star}$& 0.365 & 0.385\\ \midrule
 \multirow{1}{*}{C} & 93.9$\pm$2.5\% & 93.7$\pm$2.8\% & 93.5$\pm$1.9\% \\
 & 96.6\%& 96.9\%& 95.5\% \\ 
  & 0.385& 0.397$^{\star}$ & 0.335\\ \midrule
 \multirow{1}{*}{D} & 91.2$\pm$0.8\% & 90.5$\pm$0.2\% & 91.1$\pm$1.0\% \\
 & 92.1\%& 90.7\%& 92.2\% \\ 
  & 0.196& 0.150& 0.202$^{\star}$\\ \midrule
 \multirow{1}{*}{E} & 91.2$\pm$1.5\% & 92.8$\pm$2.2\% & 90.5$\pm$2.3\% \\
 & 92.3\%& 95.1\%& 92.4\% \\ 
  &0.211 & 0.325$^{\star}$ & 0.221   \\ \midrule
  \multirow{1}{*}{F} & 94.2$\pm$0.9\% & 93.3$\pm$0.9\% & 93.8$\pm$0.3\% \\
 & 95.3\%& 94.2\%& 94.0\% \\ 
  &0.327$^{\star}$ & 0.279& 0.273   \\ \bottomrule
  
  \multicolumn{4}{l}{$^{\star}$ marks personal best and $^{\star\star}$ overall best.}

\end{tabular}
\end{table}

In the online sessions, the performance of each subject was determined by the total time taken to correctly type the 44-character sentence (see Table~\ref{tab:online}). All subjects achieved their best results at the high interface speed and were able to complete the task within a 3.45--4.51 min time window. The subjects' average performances also demonstrate that 
they preferred the high interface speed (note that Subject A performed $n = 5$ online sessions for the 400 ms speed, while all other averages were based on $n = 6$ sessions). This was true despite the highest offline ITRs 
being achieved at the slow interface speed (cf. Table~\ref{tab:offline}), and it indicates that the speed-accuracy trade off is well exploited using an ITI=160 msec.  In addition, the practical, error-free ITRs were calculated using~(\ref{eq:itr}), and they reached values as high as 1.146 bit/sec. This bit rate corresponds to correctly typing 12.75 character/min (see supplementary movie at 
{http://www.youtube.com/user/UCIBCI}). 
 It should be noted that out of the 207.1 sec to complete the sentence, only 78.1 sec were spent on letter selection while 129 sec were spent on post-selection pauses. A similar breakdown applies to other subjects, where post-selection time constituted more than 60\% of the total time. 


\begin{table}[!htbp]
\caption{Online performances (in sec, includes the 3 sec pause after letter selection) across subjects and days. Mean performance (top row), best performance and day at which it was achieved (bottom row), and information transfer rates of the best online session (last column) are given.}\label{tab:online} 
\centering
\begin{tabular}{cllll}\toprule
\multirow{2}{*}{Subject} & \multicolumn{3}{c}{ITI (msec)} & \multirow{2}{*}{ITR (bit/sec)} \\ 
 & \multicolumn{1}{c}{400} & \multicolumn{1}{c}{240} & \multicolumn{1}{c}{160} \\ \midrule
\multirow{2}{*}{A} & 324.6$\pm$28.1 & 328.5$\pm$37.0 & 302.5$\pm$20.3\ \\
 & 302.9 (3) & 289.4 (2) & 271.7$^{\star}$ (1) & 0.873$^{\star}$\\ \midrule
 
 \multirow{2}{*}{B} &  337.9$\pm$27.7 & 304.1$\pm$48.1 & 265.0$\pm$52.5\ \\
 & 299.8 (3) & 248.9 (3) & 214.3$^{\star}$ (2) & 1.107$^{\star}$\\ \midrule
 
 \multirow{2}{*}{C} & 395.8$\pm$67.3 & 305.4$\pm$44.8 & 239.5$\pm$37.6\ \\
  & 301.8 (3) & 254.6 (3) &  207.1$^{\star\star}$ (2) & 1.146$^{\star\star}$\\ \midrule
 
 \multirow{2}{*}{D} & 495.4$\pm$61.3 & 380.7$\pm$77.0 & 306.4$\pm$86.6\ \\
 & 447.0 (1) & 275.0 (1) & 237.8$^{\star}$ (2) & 0.998$^{\star}$\\ \midrule
 
 \multirow{2}{*}{E} & 558.3$\pm$104.4 & 397.6$\pm$83.2 & 471.5$\pm$162.9\ \\
  & 471.9 (1) & 323.5 (2) &  299.7$^{\star}$ (1) & 0.792$^{\star}$\\ \midrule
  \multirow{2}{*}{F} & 465.4$\pm$118.3 & 346.2$\pm$53.4 & 254.2$\pm$14.4\ \\
  & 354.7 (3) & 263.4 (3) &  233.2$^{\star}$ (2) & 1.017$^{\star}$\\ \bottomrule
  
    \multicolumn{5}{l}{$^{\star}$ marks personal best and $^{\star\star}$ overall best.}
\end{tabular}
\end{table}

Before comparing the sustained, error-free ITRs achieved in this study to those of other EEG-based BCI systems, we make the following observations: (i) reported ITRs often exclude or simply ignore post-selection time (3 sec in the present study) from calculations~\cite{Kaperc,Meinicke2002,Serby2005}, and (ii) reported ITRs are rarely, if ever, calculated in an error-free fashion, i.e. the subjects are \textit{not} required to correct spelling errors before proceeding.
\begin{table}[!htbp]
\caption{Comparison of the best achieved information transfer rates for several EEG-based BCI studies.}\label{tab:comparison}
\begin{tabular}{ccccc}\toprule
\multicolumn{2}{c}{Study}  &ITR (bit/trial) & Trial Frequency (trial/sec) & ITR (bit/sec)\\ \midrule
\multirow{4}{*}{present} & A & 0.363 & 5.81 & 2.109 \\
& B & 0.514 & 5.82 & 2.992  \\
& C & 0.522 & 5.82 & 3.038 \\
& D & 0.422 & 5.82 & 2.455 \\ 
& E & 0.302 & 5.85 & 1.766 \\
& F & 0.415 & 5.86 & 2.434 \\\midrule
\multicolumn{2}{r}{\cite{djkrusienski:08}} & 0.039 & 5.71 & 0.224 \\ 
\multicolumn{2}{r}{\cite{gtownsend:10}} & 0.129 & 8 & 1.028 \\
\multicolumn{2}{r}{\cite{jrwolpaw:04}} & 2.373 & 0.526 & 1.249\\ 
\multicolumn{2}{r}{\cite{Guan2004}} & 0.859 & 0.463 & 0.398\\ 
\bottomrule
\end{tabular}
\end{table}
Table~\ref{tab:comparison} shows a comparison of the peak character selection ITRs achieved in the present study to those derived from other EEG-based BCI studies. The present-study ITRs were obtained from~(\ref{eq:itr}) by subtracting the total post-selection time from the personal best times reported in Table~\ref{tab:online}. While these rates were nominally achieved at ITI=160 msec (or 6.25 trial/sec), the values  reported in the middle column are somewhat lower due to real-time processing demands.
For the study in~\cite{djkrusienski:08}, the performance of the best subject (Subject A) was determined based on 11 illumination sequences (the study uses 15 sequences, but a 0\% error rate appears to be achieved after 11 sequences). This corresponds to correctly spelling at a rate of 23.1 sec ($11\times 2.1$) per character, which given a 36-character matrix and according to~(\ref{eq:itr}),  yields an ITR of 0.224 bit/sec. Also, this study used an ITI=175 msec, which is equivalent to 
5.71 trial/sec. 
Similar to~\cite{djkrusienski:08}, this corresponds to correctly spelling at a rate of 6.2 sec ($3\times 2.06$) per character. Given the same 36-character matrix and according to~(\ref{eq:itr}), this subject's result yields an ITR of 0.836 bit/sec. Since an ITI similar to that of ~\cite{djkrusienski:08} was used in this study ($\sim$172 msec), an equivalent trial frequency and ITR in bits/trial of 5.82 trials/sec and 0.144 bit/trial, respectively, can easily be calculated. 

Bit rates as high as 61.70 bit/min were reported in~\cite{gtownsend:10} (Subject 14), from which the value in Table~\ref{tab:comparison} immediately follows. The illumination frequency of 8 trials/sec readily follows from an ITI=125 msec used in the study. 
The study in~\cite{jrwolpaw:04} was not concerned with a BCI spelling task, rather it reported on BCI-controlled cursor movements to a series of 8 screen targets. The best performance (Subject D) corresponded to accuracy of 92\%, which using the formula~(\ref{eq:capacity}), yields 2.373 bit/trial. With the average duration of a trial being 1.9 sec, the ITR of 1.249 bit/trial follows readily. Finally, the study in~\cite{Guan2004} determined the speed and accuracy of two different flashing paradigms: single letter display and row-column display of flashing characters in a $6\times6$ matrix. It was determined 
that the single letter display paradigm was able to achieve higher communication speeds, 
spelling approximately 1 letter every 13 sec with a $95\%$ accuracy across all 5 participants. Thus, with this typing speed, subjects were able to spell a 42-character sentence in 546 sec, which  
corresponds to a $0.398$ bit/sec ITR using equation~(\ref{eq:itr}) while assuming error-free performances and ignoring post-selection notification times. 

\section{Discussion}


\subsection{Performance}\label{sec:p}
The above results 
dispel the common assumption regarding 
ITRs achievable by EEG-based BCIs~\cite{jrwolpaw:04, gsanthanam:06}. In particular, our system allows characters to be selected in an error-free fashion with ITRs in excess of 3 bit/sec (cf. Table~\ref{tab:comparison}), which is three times higher than the best bit rates achieved with similar
EEG spelling systems~\cite{gtownsend:10}, and  nearly three times higher than those achieved with 2-D cursor control~\cite{jrwolpaw:04}. 

The superior performance of our system can be attributed to several factors. Firstly, it is a truly single-trial system, i.e. it 
reliably classifies oddball and non-oddball stimuli after a single illumination. Other systems, such as those based on the original Farwell-Donchin paradigm~\cite{Farwell1988}, require repeated (up to 20) presentations of an oddball 
stimulus before a selection is made~\cite{djkrusienski:08,gtownsend:10}. Similar requirements are imposed in the so-called checkerboard paradigm~\cite{gtownsend:10}. 
Also, the ability to classify oddball and non-oddball stimuli on a single-trial basis with rates as high as 97.4\% (see Table~\ref{tab:offline}) is facilitated by a combination of techniques~\cite{kdas:09,Das2007,Das2008} briefly described in Section~\ref{sec:spac}. The most distinct feature of this method is that it can efficiently handle high-dimensional (480-D in the present study) spatio-temporal  data without resorting to heuristic strategies such as subsampling EEG signals~\cite{djkrusienski:08,Sellers2006} or constructing a feature set by addition/deletion of individual attributes~\cite{djkrusienski:08,gtownsend:10,Sellers2006}. This method has also been used to successfully classify other types of neurophysiologic data such as electrocorticograms (ECoG)~\cite{kdas:09a}, and in other types of BCI applications, such as asynchronous control of a virtual reality avatar~\cite{ptwang:10a,ahdo:10}, hand orthosis~\cite{ceking:11}, and functional electrical stimulator~\cite{ahdo:11}. Secondly,  biasing the illumination order of characters according to their frequencies (see Appendix~\ref{a:groc}) significantly reduces the time the subject spends waiting for the desired character to get illuminated. For example, based on 100,000 Monte Carlo trials, we estimated that the 
12 most frequent characters (see Fig.~\ref{fig:letterfrequency}) are on average illuminated within the first two groups. For comparison, if the characters had been grouped in a uniformly random fashion, both frequent (e.g. {\tt E}) and infrequent (e.g. {\tt Z}) characters would have been 
on average found in the fourth group. Thus, the above sampling procedure exploits the relatively low entropy of the English language~\cite{tmcover:91}, which in turn facilitates faster character selection.  
Similarly, the partial word completion feature prompts users to first select those letters that represent dictionary-defined completions, 
thereby bypassing stage 1 and yielding significant time savings.




\subsection{Information Transfer Rates}\label{sec:itr}
The discrepancy in per-trial ITRs between the offline (cf. Table~\ref{tab:offline}) and online performances (cf. Table~\ref{tab:comparison}) can be explained by two factors. Firstly, 
the effect of feedback and subsequent user-interface interaction (e.g. excitement, frustration), specific to online sessions, cannot be accounted for with offline data. Secondly, the 1:6 oddball-to-non-oddball ratio observed in the training procedure may be significantly higher in the online procedure. 
Since the order of character illumination depends on their frequencies, the desired characters are likely to be illuminated early in the cycle (see Appendix~\ref{a:groc}), and users are likely to select them before all 7 groups are illuminated, thereby disturbing the 1:6 ratio. To underscore this point, offline ITRs corresponding to a high interface speed (see Table~\ref{tab:offline}) were recalculated according to~(\ref{eq:mi}) with the assumed 1:3 oddball-to-non-oddball ratio,  and values between 0.274 and 0.524 bit/trial were obtained, which are remarkably close to the ITRs achieved online (see Table~\ref{tab:comparison}). Therefore, the mutual information formula~(\ref{eq:mi}) provides a reasonably accurate estimate of ITRs achievable online. 

Based on the above, it follows that the formula~\cite{Pierce1980}:
\begin{equation}\label{eq:capacity}
I(\mathrm{in,out})= \log_2C + p_c\log_2 p_c + p_{\varepsilon}\log_2\left(\frac{p_{\varepsilon}}{C-1}\right)
\end{equation}
frequently used to express ITRs in BCI studies~\cite{Wolpaw2002,Nam2010,Wolpaw2000}, 
is not adequate for BCIs based on the oddball paradigm. First note that for a two-class system ($C=2$), the expression~(\ref{eq:capacity}) reduces to $I(\mathrm{in,out})=1+p_c\log_2 p_c + p_{\varepsilon}\log_2 p_{\varepsilon}$, which represents the capacity (the maximum achievable ITR) of a binary \textit{symmetric} channel~\cite{tmcover:91}. To achieve this upper limit, in addition to being symmetric, the BCI communication channel must maintain equal prior probabilities of oddball and non-oddball trials, i.e. $p(o)= p(e)$. This, however, contradicts the very definition of the oddball paradigm, where by design we must have $p(o)\ll p(e)$. Similarly, for a chance level performance, we have $p_c=p(e)$ and $p_{\varepsilon}=p(o)$ (see Appendix~\ref{a:itrc}), and so it follows from~(\ref{eq:capacity}) (assuming the standard $p(o)$ to $p(e)$ ratio) that $I(\mathrm{in,out})=0.408$ bit/trial, which presents an obvious contradiction. On the other hand, our analysis correctly demonstrates (Appendix~\ref{a:itrc}) that subjects with performances $p_c\le p(e)$ are not able to spell, since $I(\mathrm{in,out})=0$. Depending on the exact oddball-to-non-oddball ratio, offline performances substantially greater than $50\%$ may be necessary for successful online spelling.    

For an asymmetric communication channel (Appendix~\ref{a:itrc}), the probabilities $p_{\varepsilon}$ and $p_c$ cannot be unequivocally linked to the mutual information, i.e.  the confusion matrix probabilities must be used explicitly. 
If these are not available, a lower bound  based on the Fano inequality~\cite{tmcover:91} may be used
\begin{equation}
\label{eq:fano}
I(\mathrm{in,out}) \ge H(\mathrm{in})+p_c\log_2 p_c + p_{\varepsilon}\log_2(p_{\varepsilon}) 
\end{equation}
where $H(\mathrm{in})=-\left[p({o})\log_2p({o})+ p({e})\log_2p({e})\right]$ (similar to~(\ref{eq:hout})). 
Likewise, an upper bound on the mutual information may be derived from the Hellman-Raviv inequality~\cite{znenadic:07,mhellman:70}.

\subsection{Improvements} 
While it has achieved unprecedented, error-free, online typing rates, our BCI-speller has not been optimized. For example, as the users underwent multiple experiments, 
they became familiar with the character layout, and felt that further reduction of post-selection pause (e.g. from 3 to 2 sec) would not compromise the spelling accuracy. This step alone would have reduced the total spelling times (see Table~\ref{tab:online}) by at least 43 sec, and increased the practical, error-free ITRs by at least 25\%. 
Furthermore, addition of a full word completion feature similar to current text-messaging systems 
could further significantly increase the practical bit rates. Implementation of these improvements is straightforward, 
although some user training may be required. 

Through a more elaborate procedure, 
the performance of the BCI-speller with various omission and false alarm rates could be simulated. 
This would allow the costs $\lambda_{\mathrm{OM}}$ and $\lambda_{\mathrm{FA}}$ (see Section~\ref{sec:spac}) to be evaluated more objectively and the classification threshold in~(\ref{eq:bayes}) to be chosen optimally. The optimal threshold is likely to be interface-speed dependent, as the two costs scale differently with the interface speed due to the fact that each false alarm incurs a constant cost associated with the post-selection pause. Another potential improvement could be achieved if in addition to training on stage-1 data (see Section~\ref{sec:ep}), the parameters of CPCA, AIDA and the Bayesian classifier were estimated from stage-2 data. Together with more objectively estimated $\lambda_{\mathrm{OM}}$ and $\lambda_{\mathrm{FA}}$, this would allow the stage-2 
threshold to be set more accurately (as opposed to the empirically chosen $\theta=0.5$).  
In addition, 
fine tuning of ITI  
and increasing the number of channels, especially over the parietal lobe~\cite{djkrusienski:08}, could conceivably improve the ITRs even further. Finally, optimization of 
luminance~\cite{ktakano:09}, background/foreground color, and character size and spacing~\cite{Salvaris2009}, may lead to further improvements. 

\subsection{Conclusion}
By exploiting basic concepts from pattern recognition theory and information theory, the presented EEG-based BCI communication system allows for error-free selection of characters with sustained, online bit rates that are several-fold higher than those that have been achieved with similar BCI systems. More importantly, these results disprove the common assumption that ITRs of EEG-based BCI systems are limited to $\sim$1 bit/sec~\cite{Wolpaw2002,gsanthanam:06}. Since the parameters of the present system have not been optimized, we hypothesize that further substantial improvements of both character-selection and practical ITRs can be achieved. Many of these improvements are straightforward, while others may require some user training. These results may have significant implications on the viability and adoption of EEG-based BCIs in both clinical and non-clinical applications. They also offer compelling evidence for further development of state-of-the-art statistical signal processing and pattern recognition methods aimed at the single-trial processing and analysis of high-dimensional EEG data.

\appendices
\section{Group Randomization of Characters}\label{a:groc}
\begin{figure}[!htbp]
        \centering
                \includegraphics[width=1.0\columnwidth]{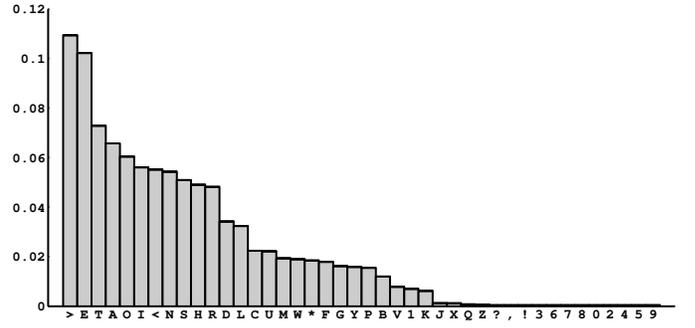}
        \caption{The frequency distribution of the characters in the BCI speller.}
        \label{fig:letterfrequency}
\end{figure}

During BCI operation, 
the characters are highlighted in a random order, biased to the frequency of the English language alphabet, digits, and punctuations (see Fig.~\ref{fig:letterfrequency})~\cite{eslee:99, Lewand2000}.
Before each stage-1 cycle, the order of characters is re-randomized in an iterative process by using the inverse sampling theorem~\cite{ldevroye:86}. To this end, a cumulative distribution function (CDF) $F_X(x)$, $x=\{{\tt >}, {\tt E}, {\tt T},  \cdots\}$, is  calculated by integrating the character histogram. 
The inverse sampling theorem states that
if $Y$ is a uniformly distributed random variable, then the CDF of 
$X\triangleq F_X^{-1}(Y)$ is precisely $F_X$, and so uniform
distribution can be mapped into any distribution whose 
$F_X$ is known.  
To order characters according to their frequencies, at each iteration a uniformly sampled random number, $y^{\star}=\mathcal{U}[0,1]$, 
was mapped according to  $x^{\star}=F_X^{-1}(y^{\star})$, and the character corresponding to $x^{\star}$ was drawn without replacement.
Since $F_X$ is discontinuous, the mapping $F_X^{-1}$ is implemented as a lookup table. 
This procedure is then iterated until all 42 characters have been drawn. The characters are subsequently organized into 7 groups of 6 characters according to the order in which they were drawn.  
Using this approach, the more frequently used characters, e.g. \{{\tt >}, {\tt E}, {\tt T}, {\tt A}, {\tt O}, {\tt I}\} are likely to be highlighted earlier in a cycle. For example, Monte Carlo simulations show that the average number of illuminations necessary to highlight  {\tt >} (the most frequent character) is only 1.6 (out of 7). 
Note, however, that since the above algorithm is stochastic, the 
order of illumination 
varies over cycles, which prevents the formation of predictable spatio-temporal illumination patterns that are known to weaken the P300 response.
Finally, in stage 2, the 
characters are illuminated one-by-one in the same biased random order as in stage 1.


\section{Classifier Design}\label{a:cd}
In Bayesian decision theory~\cite{roduda:01}, the loss associated with each incorrect decision is expressed by:
\begin{eqnarray}\label{eq:cond_risk}
\mathcal{R}(o\,|\,F^{\star}) &=& \lambda_{\mathrm{FA}} p(e\,|\,F^{\star})\\
\mathcal{R}(e\,|\,F^{\star}) &=& \lambda_{\mathrm{OM}} p(o\,|\,F^{\star})
\end{eqnarray}
where $\mathcal{R}(o\,|\,F^{\star})$ is the (conditional) risk of classifying observation $F^{\star}$ as oddball. It is the product of the probability that the identity of $F^{\star}$ is not oddball, i.e. $p(e\,|\,F^{\star})$, and the cost of such decision is $\lambda_{\mathrm{FA}}$. The risk associated with classifying $F^{\star}$ as non-oddball, $\mathcal{R}(e\,|\,F^{\star})$, is defined in a similar manner. The overall risk is then defined as:
\begin{equation}
\label{eq:risk}
\mathcal{R}=\int_{\Omega} \left[\mathcal{R}(o\,|\,F)+\mathcal{R}(e\,|\,F)\right] f(F)\,dF
\end{equation}
where $f(F)$ is the probability density function (PDF) of features, $F$. The minimum value of~(\ref{eq:risk}) is known as the Bayes risk, and the decision rule that accomplishes this optimum is:
\begin{equation}
\label{eq:decision}
\mathcal{R}(e\,|\,F^{\star})
\begin{array}{c}
o\\[-0.075in]
>\\[-0.075in]
<\\[-0.075in]
e       
\end{array}
\mathcal{R}(o\,|\,F^{\star})\qquad \forall F^{\star} \in \Omega
\end{equation}
i.e. make a decision that carries the smaller risk. The Bayesian classifier~(\ref{eq:bayes}) then follows readily from~(\ref{eq:decision}).

To estimate 
its parameters, the Bayes theorem is used:
\begin{equation*}
p(o\,|\,F) = \frac{f(F\,|\,o)p(o)}{f(F)}
\end{equation*}
where $p(o)$ is the prior probability of oddball trials and $f(F\,|\,o)$ is the conditional PDF of features under the oddball class. Similar expression is derived for $p(e\,|\,F)$. The Bayesian classifier~(\ref{eq:bayes}) is then implemented as a likelihood ratio test: 
\begin{equation}
\label{eq:lrt}
\frac{f(F^{\star}\,|\,o)}{f(F^{\star}\,|\,e)}
\begin{array}{c}
o\\[-0.075in]
>\\[-0.075in]
<\\[-0.075in]
e
\end{array}
\frac{\lambda_{\mathrm{FA}}}{\lambda_{\mathrm{OM}}}
\frac{p(e)}{p(o)}
\end{equation}
The parameters of the linear Bayesian classifier are estimated by assuming features that are conditionally Gaussian, i.e. $F\,|\,o\sim\mathcal{N}(\hat{\mu}_{o},\hat{\sigma}^2)$ and $F\,|\,e\sim\mathcal{N}(\hat{\mu}_{e},\hat{\sigma}^2)$, where $\hat{\mu}_{o}$ and $\hat{\mu}_e$ are conditional sample means of features under oddball and non-oddball classes, respectively, and $\hat{\sigma}^2$ is the (unconditional) sample variance of features~\cite{roduda:01}. 

We conclude by noting that if  $\lambda_{\mathrm{FA}}=\lambda_{\mathrm{OM}}$ and the features are non-informative, i.e. $f(F\,|\,o)=f(F\,|\,e)$, the classifier's decision~(\ref{eq:lrt}) hinges solely upon the ratio of $p(e)$ and $p(o)$. Since oddball experimental paradigms imply $p(e)>p(o)$, the Bayesian classifier~(\ref{eq:lrt}) will always pick the non-oddball class in this case. This decision rule defines the chance-level performance of the Bayesian classifier (see Appendix~\ref{a:itrc}). 

\section{ITR Calculation}\label{a:itrc}
\begin{figure}[!htbp]
        \centering
                \includegraphics[width=1.75in]{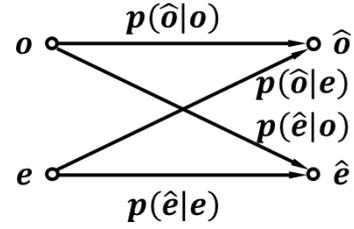}
        \caption{BCI system as a noisy communication channel.}
        \label{fig:comm_chan}
\end{figure}

Fig.~\ref{fig:comm_chan} shows the schematic of an asymmetric binary communication channel with inputs: $o$ (intent to select the highlighted character--oddball) and $e$ (intent not to select the highlighted character--non-oddball), and outputs: $\hat{o}$ (oddball detected) and $\hat{e}$ (non-oddball detected). The transition probabilities between inputs and outputs can be estimated by the 10-fold CV (see Section~\ref{sec:spac}) and comparing the true and decoded identities of test trials.   
This procedure yields a confusion matrix:
\begin{equation}
\label{eq:cf}
C_M = \left[
\begin{array}{cc}
p(\hat{o}\,|\,o) & p(\hat{e}\,|\,o)\\
p(\hat{o}\,|\,e) & p(\hat{e}\,|\,e)
\end{array}
\right]
\end{equation}
where $p(\hat{o}\,|\,o)$ and $p(\hat{e}\,|\,e)$ are the fraction of correctly decoded oddball (non-oddball) trials, respectively.
Note that $p(\hat{e}\,|\,o)=1 - p(\hat{o}\,|\,o)$ represents the probability of omission and  
$p(\hat{o}\,|\,e)=1 - p(\hat{e}\,|\,e)$ is the probability of a false alarm. The probability of error (misclassification) is then defined as: $p_{\varepsilon}=p(\hat{e}\,|\,o)p(o) + p(\hat{o}\,|\,e)p(e)$. 
Finally, the probability of correct classification is defined as: $p_c = 1 - p_{\varepsilon}$.

By the law of total probability, we have:
\begin{eqnarray}
\nonumber 
p(\hat{o})= p(\hat{o}\,|\,o)p(o) + p(\hat{o}\,|\,e)p(e)\\ \label{eq:priors}
p(\hat{e})= p(\hat{e}\,|\,o)p(o) + p(\hat{e}\,|\,e)p(e)
\end{eqnarray}
where $p(\hat{o})$ and  $p(\hat{e})$ are simply the (unconditional) probabilities of decoded trials. Since the output of the communication channel takes the values $\hat{o}$ and $\hat{e}$, 
then its unconditional entropy~\cite{tmcover:91} is given by:
\begin{equation}
\label{eq:hout}
H(\mathrm{out})=-\left[p(\hat{o})\log_2p(\hat{o})+ p(\hat{e})\log_2p(\hat{e})\right]
\end{equation} 
To complete the calculation of the mutual information~(\ref{eq:mi}), we estimate the conditional entropy, $H(\mathrm{out}\,|\,\mathrm{in})$, as in~\cite{tmcover:91}:
\begin{eqnarray}\nonumber
H(\mathrm{out}\,|\,\mathrm{in})=H(\mathrm{out}\,|\,\mathrm{in}=o)p(o)+H(\mathrm{out}\,|\,\mathrm{in}=e)p(e)\\ \nonumber
= -\left[p(\hat{o}\,|\,o)\log_2p(\hat{o}\,|\,o)+ p(\hat{e}\,|\,o)\log_2p(\hat{e}\,|\,o)\right]p(o)\\ \label{eq:hcond}
 -\left[p(\hat{o}\,|\,e)\log_2p(\hat{o}\,|\,e)+ p(\hat{e}\,|\,e)\log_2p(\hat{e}\,|\,e)\right]p(e) 
\end{eqnarray}
Thus, the mutual information can be calculated by subtracting~(\ref{eq:hcond}) from~(\ref{eq:hout}). 
It can also be shown that unless a communication channel is symmetric, i.e. $p(\hat{e}\,|\,o)=p(\hat{o}\,|\,e)$, $I(\mathrm{in},\mathrm{out})$ \textit{cannot} be expressed as a function $p_\varepsilon$ and $p_c$.

If observed features, $F^{\star}$, carry no class-relevant information, the Bayesian classifier~(\ref{eq:lrt}) will assign them to the more numerous class (non-oddball), regardless of their class identities (see Appendix~\ref{a:cd}). The confusion matrix entries then become:  $p(\hat{o}\,|\,o)=p(\hat{o}\,|\,e)=0$ and $p(\hat{e}\,|\,o)=p(\hat{e}\,|\,e)=1$, and the chance-level performance of the Bayesian classifier is: $p_{\varepsilon}=p(o)$ and $p_c=p(e)$. Assuming the oddball to non-oddball ratio of 1:6, we arrive at the chance performance: $p_c=85.71\%$. 
Under this condition, it readily follows from~(\ref{eq:hout}) 
and~(\ref{eq:hcond}) (after using l'H\^{o}pital's rule) that $H(\mathrm{out})=0$ and $H(\mathrm{out}\,|\,\mathrm{in})=0$. Therefore, the chance-level performance indeed yields $I(\mathrm{in},\mathrm{out})=0$.

In the case of the perfect classifier, we have  $p(\hat{o}\,|\,o)=p(\hat{e}\,|\,e)=1$, and so from~(\ref{eq:priors}) we obtain $p(\hat{o})=p(o)$ and $p(\hat{e})=p(e)$. It follows immediately that $H(\mathrm{out}\,|\,\mathrm{in})=0$, since the output is no longer considered random, once the input is known. 
Assuming the same 1:6 oddball-to-non-oddball ratio, we obtain: $I(\mathrm{out},\mathrm{in})=H(\mathrm{out})=0.592$ bit/transmission, which is the theoretical maximum of this 
channel. 

\bibliographystyle{unsrt}
\bibliography{../IEEEabrv,../p300}

\begin{thebibliography}{10}

\bibitem{nbirbaumer:99}
N.~Birbaumer, N.~Ghanayim, T.~Hinterberger, I.~Iversen, B.~Kotchoubey,
  A.~K\"{u}bler, J.~Perelmouter, E.~Taub, and H.~Flor.
\newblock A spelling device for the paralysed.
\newblock {\em Nature}, 398(6725):297--298, 1999.

\bibitem{Wolpaw2002}
J.R. Wolpaw, N.~Birbaumer, D.J. McFarland, G.~Pfurtscheller, and T.M. Vaughan.
\newblock Brain-computer interfaces for communication and control.
\newblock {\em Clin Neurophysiol}, 113(6):767--791, 2002.

\bibitem{jrwolpaw:04}
J.R. Wolpaw and D.J. McFarland.
\newblock Control of a two-dimensional movement signal by a noninvasive
  brain-computer interface in humans.
\newblock {\em Proc Natl Acad Sci USA}, 101(51):17849--17854, 2004.

\bibitem{djmcfarland:10}
D.J. McFarland, W.A. Sarnacki, and J.R. Wolpaw.
\newblock Electroencephalographic ({EEG}) control of three-dimensional
  movement.
\newblock {\em J Neural Eng}, 7(3):036007, 2010.

\bibitem{rleeb:07}
R.~Leeb, D.~Friedman, G.R. M\"{u}ller-Putz, R.~Scherer, M.~Slater, and
  G.~Pfurtscheller.
\newblock Self-paced (asynchronous) {BCI} control of a wheelchair in virtual
  environments: a case study with a tetraplegic.
\newblock {\em Comput Intell Neurosci}, page 79642, 2007.

\bibitem{ptwang:10a}
P.T. Wang, C.E. King, L.A. Chui, Z.~Nenadic, and A.H. Do.
\newblock {BCI} controlled walking simulator for a {BCI} driven {FES} device.
\newblock In {\em Proc. of RESNA Annual Conference}, Las Vegas, Nevada, 2010.

\bibitem{ahdo:11}
A.H. Do, P.T. Wang, A.~Abiri, C.E. King, and Z.~Nenadic.
\newblock Brain-computer interface controlled functional electrical stimulation
  system for ankle movement.
\newblock {\em J Neuroeng Rehabil}, 8:49, 2011.

\bibitem{grmuellerputz:05}
G.R. M\"{u}ller-Putz, R.~Scherer, G.~Pfurtscheller, and R.~Rupp.
\newblock {EEG}-based neuroprosthesis control: a step towards clinical
  practice.
\newblock {\em Neurosci Lett}, 382(1-2):169--174, 2005.

\bibitem{gpfurtscheller:03}
G.~Pfurtscheller, G.R. M\"{u}ller, J.~Pfurtscheller, H.J. Gerner, and R.~Rupp.
\newblock `{T}hought'--control of functional electrical stimulation to restore
  hand grasp in a patient with tetraplegia.
\newblock {\em Neurosci Lett}, 351(1):33--36, 2003.

\bibitem{Farwell1988}
L.A. Farwell and E.~Donchin.
\newblock Talking off the top of your head: toward a mental prosthesis
  utilizing event-related brain potentials.
\newblock {\em Electroencephalogr Clin Neurophysiol}, 70(6):510--523, 1988.

\bibitem{ssutton:65}
S.~Sutton, M.~Braren, J.~Zubin, and E.R. John.
\newblock Evoked-potential correlates of stimulus uncertainty.
\newblock {\em Science}, 150(700):1187--1188, 1965.

\bibitem{kcsquires:77}
K.C. Squires, S.~Petuchowski, C.~Wickens, and E.~Donchin.
\newblock The effects of stimulus sequence on event related potentials: a
  comparison of visual and auditory sequences.
\newblock {\em Perception \& Psychophysics}, 22(1):31--40, 1977.

\bibitem{djkrusienski:08}
D.J. Krusienski, E.W. Sellers, D.J. McFarland, T.M. Vaughan, and J.R. Wolpaw.
\newblock Toward enhanced {P}300 speller performance.
\newblock {\em J Neurosci Methods}, 167(1):15--21, 2008.

\bibitem{gtownsend:10}
G.~Townsend, B.K. LaPallo, C.B. Boulay, D.J. Krusienski, G.E. Frye, C.K.
  Hauser, N.E. Schwartz, T.M. Vaughan, J.R. Wolpaw, and E.W. Sellers.
\newblock A novel {P}300-based brain-computer interface stimulus presentation
  paradigm: moving beyond rows and columns.
\newblock {\em Clin Neurophysiol}, 121(7):1109--1120, 2010.

\bibitem{tehutchinson:89}
T.E. Hutchinson, K.P. White, W.N. Martin, K.C. Reichert, and L.A. Frey.
\newblock Human-computer interaction using eye-gaze input.
\newblock {\em {IEEE} Trans. Syst., Man, Cybern.}, 19(6):1527--1534, 1989.

\bibitem{sthorpe:96}
S.~Thorpe, D.~Fize, and C.~Marlot.
\newblock Speed of processing in the human visual system.
\newblock {\em Nature}, 381(6582):520--522, 1996.

\bibitem{kdas:09}
K.~Das and Z.~Nenadic.
\newblock An efficient discriminant-based solution for small sample size
  problem.
\newblock {\em Pattern Recogn}, 42(5):857--866, 2009.

\bibitem{Das2007}
K.~Das, S.~Osechinskiy, and Z.~Nenadic.
\newblock A classwise {PCA}-based recognition of neural data for brain-computer
  interfaces.
\newblock In {\em Proc. 29th Annual Int. Conf. of the IEEE Engineering in
  Medicine and Biology Society}, pages 6520--6523, 2007.

\bibitem{Das2008}
K.~Das and Z.~Nenadic.
\newblock Approximate information discriminant analysis: a computationally
  simple heteroscedastic feature extraction technique.
\newblock {\em Pattern Recogn}, 41(5):1548--1557, 2008.

\bibitem{stroweis:00}
S.T. Roweis and L.K. Saul.
\newblock Nonlinear dimensionality reduction by locally linear embedding.
\newblock {\em Science}, 290(5500):2323--2326, 2000.

\bibitem{znenadic:07}
Z.~Nenadic.
\newblock Information discriminant analysis: Feature extraction with an
  information-theoretic objective.
\newblock {\em {IEEE} Trans. Pattern Anal. Mach. Intell.}, 29(8):1394--1407,
  2007.

\bibitem{ktorkkola:03}
K.~Torkkola.
\newblock Feature extraction by non-paramatric mutual information maximization.
\newblock {\em J. Mach. Learn. Res.}, 3:1415--1438, 2003.

\bibitem{roduda:01}
R.O. Duda, P.E. Hart, and D.G. Stork.
\newblock {\em Pattern Classification}.
\newblock Wiley-Interscience, New York, 2001.

\bibitem{znenadic:05a}
Z.~Nenadic and J.W. Burdick.
\newblock Spike detection using the continuous wavelet transform.
\newblock {\em IEEE T. Biomed. Eng.}, 52(1):74--87, 2005.

\bibitem{rkohavi:95}
R.~Kohavi.
\newblock A study of cross-validation and bootstrap for accuracy estimation and
  model selection.
\newblock In {\em Int. Joint C. Art. Int.}, pages 1137--1145, 1995.

\bibitem{tmcover:91}
T.M. Cover and J.A. Thomas.
\newblock {\em Elements of Information Theory.}
\newblock Wiley Interscience, New York, 1991.

\bibitem{ewsellers:07}
E.W. Sellers, D.J. Krusienski, D.J. McFarland, and J.R. Wolpaw.
\newblock Non-invasive brain-computer interface research at the {W}adsworth
  {C}enter.
\newblock In G.~Dornhege, J.R. Mill\'{a}n, T.~Hinterberger, D.J. McFarland, and
  K.-R. M\"{u}ller, editors, {\em Toward Brain-Computer Interfacing}, pages
  31--42. The {MIT} Press, 2007.

\bibitem{plnunez:06}
P.L. Nunez and R.~Srinivasan.
\newblock {\em Electric Fields of the Brain: The Neurophysics of {EEG}}.
\newblock Oxford University Press, New York, 2nd edition, 2006.

\bibitem{clgonsalvez:02}
C.L. Gonsalvez and J.~Polich.
\newblock P300 amplitude is determined by target-to-target interval.
\newblock {\em Psychophysiology}, 39(3):388--396, 2002.

\bibitem{Kaperc}
M.~Kaper, P.~Meinicke, U.~Grossekathoefer, T.~Lingner, and H.~Ritter.
\newblock {BCI} competition 2003-data set {IIb}: support vector machines for
  the {P300} speller paradigm.
\newblock {\em {IEEE} Trans. Biomed. Eng.}, 51(6):1073--1076.

\bibitem{Meinicke2002}
P.~Meinicke, M.~Kaper, F.~Hoppe, M.~Heumann, and H.~Ritter.
\newblock Improving transfer rates in brain computer interfacing: A case study.
\newblock In {\em NIPS}, pages 1107--1114, 2002.

\bibitem{Serby2005}
H.~Serby, E.~Yom-Tov, and G.F. Inbar.
\newblock An improved {P}300-based brain-computer interface.
\newblock {\em {IEEE} Trans. Neural Syst. Rehabil. Eng.}, 13(1):89--98, 2005.

\bibitem{Guan2004}
C.~Guan, M.~Thulasidas, and J.~Wu.
\newblock High performance {P300} speller for brain-computer interface.
\newblock In {\em IEEE International Workshop on Biomedical Circuits and
  Systems}. IEEE International Workshop, 2004.

\bibitem{gsanthanam:06}
G.~Santhanam, S.I. Ryu, B.M. Yu, A.~Afshar, and K.V. Shenoy.
\newblock A high-performance brain-computer interface.
\newblock {\em Nature}, 442(7099):195--198, 2006.

\bibitem{Sellers2006}
E.W. Sellers, D.J. Krusienski, D.J. McFarland, T.M. Vaughan, and J.R. Wolpaw.
\newblock A {P300} event-related potential brain-computer interface ({BCI}):
  the effects of matrix size and inter stimulus interval on performance.
\newblock {\em Biol Psychol}, 73(3):242--252, 2006.

\bibitem{kdas:09a}
K.~Das, D.S. Rizzuto, and Z.~Nenadic.
\newblock Mental state estimation for brain--computer interfaces.
\newblock {\em {IEEE} Trans. Biomed. Eng.}, 56(8):2114--2122, 2009.

\bibitem{ahdo:10}
A.H. Do, P.T. Wang, C.E. King, L.A. Chui, and Z.~Nenadic.
\newblock Asynchronous {BCI} control of a walking simulator.
\newblock In {\em the Fourth International {BCI} Meeting}, Asilomar, CA, June,
  2010.

\bibitem{ceking:11}
C.E. King, P.T. Wang, M.~Mizuta, D.J. Reinkensmeyer, A.H. Do, S.~Moromugi, and
  Z.~Nenadic.
\newblock Noninvasive brain-computer interface driven hand orthosis.
\newblock In {\em Proc. 33rd Annual Int. Conf. of the IEEE Engineering in
  Medicine and Biology Society}, pages 5786--5789, 2011.

\bibitem{Pierce1980}
J.R. Pierce.
\newblock {\em An introduction to information theory: symbols, signals, and
  noise}.
\newblock Dover Publications, Inc., New York, 2nd edition, 1980.

\bibitem{Nam2010}
C.S. Nam, Y.~Li, and S.~Johnson.
\newblock Evaluation of a {P}300-based brain-computer interface in real-world
  contexts.
\newblock {\em Int J Hum-Comput Int}, 26(6):621--637, 2010.

\bibitem{Wolpaw2000}
J.R. Wolpaw, N.~Birbaumer, W.J. Heetderks, D.J. McFarland, P.H. Peckham,
  G.~Schalk, E.~Donchin, L.A. Quatrano, C.J. Robinson, and T.M. Vaughan.
\newblock Brain-computer interface technology: a review of the first
  international meeting.
\newblock {\em {IEEE} Trans. Rehabil. Eng.}, 8(2):164--173, 2000.

\bibitem{mhellman:70}
M.~Hellman and J.~Raviv.
\newblock Probability of error, equivocation, and the {C}hernoff bound.
\newblock {\em {IEEE} Trans. Inf. Theory}, 16(4):368--372, 1970.

\bibitem{ktakano:09}
K.~Takano, T.~Komatsu, N.~Hata, Y.~Nakajima, and K.~Kansaku.
\newblock Visual stimuli for the {P300} brain-computer interface: a comparison
  of white/gray and green/blue flicker matrices.
\newblock {\em Clin Neurophysiol}, 120(8):1562--1566, 2009.

\bibitem{Salvaris2009}
M.~Salvaris and F.~Sepulveda.
\newblock Visual modifications on the {P300} speller {BCI} paradigm.
\newblock {\em J Neural Eng}, 6(4):046011, 2009.

\bibitem{eslee:99}
E.~Stewart Lee.
\newblock {\em Essays about computer security}, page 181.
\newblock University of Cambridge Computer Laboratory, 1999.

\bibitem{Lewand2000}
R.~Lewand.
\newblock {\em Cryptological mathematics}, page~36.
\newblock The Mathematical Association of America, 2000.

\bibitem{ldevroye:86}
L.~Devroye.
\newblock {\em Non-Uniform Random Variate Generation}, chapter~2, pages 27--39.
\newblock Springer-Verlag, New York, 1986.

\end{thebibliography}
\end{document}